\PassOptionsToPackage{unicode}{hyperref}
\PassOptionsToPackage{hyphens}{url}
\PassOptionsToPackage{dvipsnames,svgnames,x11names}{xcolor}
\documentclass[
]{article}
\usepackage{xcolor}
\usepackage{amsmath,amssymb}
\usepackage{iftex}
\ifPDFTeX
  \usepackage[T1]{fontenc}
  \usepackage[utf8]{inputenc}
  \usepackage{textcomp} % provide euro and other symbols
\else % if luatex or xetex
  \usepackage{unicode-math} % this also loads fontspec
  \defaultfontfeatures{Scale=MatchLowercase}
  \defaultfontfeatures[\rmfamily]{Ligatures=TeX,Scale=1}
\fi
\usepackage{lmodern}
\ifPDFTeX\else
\fi
\IfFileExists{upquote.sty}{\usepackage{upquote}}{}
\IfFileExists{microtype.sty}{% use microtype if available
  \usepackage[]{microtype}
  \UseMicrotypeSet[protrusion]{basicmath} % disable protrusion for tt fonts
}{}
\makeatletter
\@ifundefined{KOMAClassName}{% if non-KOMA class
  \IfFileExists{parskip.sty}{%
    \usepackage{parskip}
  }{% else
    \setlength{\parindent}{0pt}
    \setlength{\parskip}{6pt plus 2pt minus 1pt}}
}{% if KOMA class
  \KOMAoptions{parskip=half}}
\makeatother
\makeatletter
\ifx\paragraph\undefined\else
  \let\oldparagraph\paragraph
  \renewcommand{\paragraph}{
    \@ifstar
      \xxxParagraphStar
      \xxxParagraphNoStar
  }
  \newcommand{\xxxParagraphStar}[1]{\oldparagraph*{#1}\mbox{}}
  \newcommand{\xxxParagraphNoStar}[1]{\oldparagraph{#1}\mbox{}}
\fi
\ifx\subparagraph\undefined\else
  \let\oldsubparagraph\subparagraph
  \renewcommand{\subparagraph}{
    \@ifstar
      \xxxSubParagraphStar
      \xxxSubParagraphNoStar
  }
  \newcommand{\xxxSubParagraphStar}[1]{\oldsubparagraph*{#1}\mbox{}}
  \newcommand{\xxxSubParagraphNoStar}[1]{\oldsubparagraph{#1}\mbox{}}
\fi
\makeatother

\usepackage{longtable,booktabs,array}
\usepackage{calc} % for calculating minipage widths
\usepackage{etoolbox}
\makeatletter
\patchcmd\longtable{\par}{\if@noskipsec\mbox{}\fi\par}{}{}
\makeatother
\IfFileExists{footnotehyper.sty}{\usepackage{footnotehyper}}{\usepackage{footnote}}
\makesavenoteenv{longtable}
\usepackage{graphicx}
\makeatletter
\newsavebox\pandoc@box
\newcommand*\pandocbounded[1]{% scales image to fit in text height/width
  \sbox\pandoc@box{#1}%
  \Gscale@div\@tempa{\textheight}{\dimexpr\ht\pandoc@box+\dp\pandoc@box\relax}%
  \Gscale@div\@tempb{\linewidth}{\wd\pandoc@box}%
  \ifdim\@tempb\p@<\@tempa\p@\let\@tempa\@tempb\fi% select the smaller of both
  \ifdim\@tempa\p@<\p@\scalebox{\@tempa}{\usebox\pandoc@box}%
  \else\usebox{\pandoc@box}%
  \fi%
}
\def\fps@figure{htbp}
\makeatother

\providecommand{\tightlist}{%
  \setlength{\itemsep}{0pt}\setlength{\parskip}{0pt}}

\usepackage[]{natbib}
\usepackage{moreverb,url}

\usepackage[colorlinks,bookmarksopen,bookmarksnumbered,citecolor=red,urlcolor=red]{hyperref}

\usepackage{booktabs}
\usepackage{appendix}

\newcommand\BibTeX{{\rmfamily B\kern-.05em \textsc{i\kern-.025em b}\kern-.08em
T\kern-.1667em\lower.7ex\hbox{E}\kern-.125emX}}

\newcommand\numberthis{\addtocounter{equation}{1}\tag{\theequation}}

\usepackage{subcaption}
\usepackage{arxiv}
\usepackage{orcidlink}
\usepackage{amsmath}
\usepackage[T1]{fontenc}
\usepackage{rotating}
\makeatletter
\@ifpackageloaded{caption}{}{\usepackage{caption}}
\AtBeginDocument{%
\ifdefined\contentsname
  \renewcommand*\contentsname{Table of contents}
\else
  \newcommand\contentsname{Table of contents}
\fi
\ifdefined\listfigurename
  \renewcommand*\listfigurename{List of Figures}
\else
  \newcommand\listfigurename{List of Figures}
\fi
\ifdefined\listtablename
  \renewcommand*\listtablename{List of Tables}
\else
  \newcommand\listtablename{List of Tables}
\fi
\ifdefined\figurename
  \renewcommand*\figurename{Figure}
\else
  \newcommand\figurename{Figure}
\fi
\ifdefined\tablename
  \renewcommand*\tablename{Table}
\else
  \newcommand\tablename{Table}
\fi
}
\@ifpackageloaded{float}{}{\usepackage{float}}
\floatstyle{ruled}
\@ifundefined{c@chapter}{\newfloat{codelisting}{h}{lop}}{\newfloat{codelisting}{h}{lop}[chapter]}
\floatname{codelisting}{Listing}

\makeatother
\makeatletter
\@ifpackageloaded{caption}{}{\usepackage{caption}}
\@ifpackageloaded{subcaption}{}{\usepackage{subcaption}}
\makeatother
\usepackage{bookmark}
\IfFileExists{xurl.sty}{\usepackage{xurl}}{} % add URL line breaks if available
\hypersetup{
  pdftitle={Interrupted Time Series Analysis of Count Data with Nuisance Interruptions},
  pdfauthor={Benjamin Stockton; Linda G. Kahn; Shilpi S. Mehta-Lee; Erinn M. Hade},
  pdfkeywords={Causal inference, Bayesian data analysis, Stacking, Model
combination, Electronic health records, Maternal health},
  colorlinks=true,
  linkcolor={blue},
  filecolor={Maroon},
  citecolor={Blue},
  urlcolor={Blue},
  pdfcreator={LaTeX via pandoc}}

\usepackage{setspace}
\renewcommand{\today}{2026-07-15}
\newcommand{\runninghead}{A Preprint }
\renewcommand{\runninghead}{A Preprint }
\title{Interrupted Time Series Analysis of Count Data with Nuisance
Interruptions}
\def\asep{\\\\\\ } % default: all authors on same column
\author{\textbf{Benjamin
Stockton}~\orcidlink{0000-0002-3820-5293}\\Department of Population
Health\\New York University Grossman School of Medicine\\New
York,\ 10016\\\href{mailto:benjamin.stockton@nyulangone.org}{benjamin.stockton@nyulangone.org}\asep\textbf{Linda
G. Kahn}\\Department of Population Health\\New York University Grossman
School of Medicine\\New York,\ 10016\\Department of Pediatrics\\New York
University Grossman School of Medicine\\New
York,\ 10016\\\asep\textbf{Shilpi S. Mehta-Lee}\\Department of
Obstetrics and Gynecology\\New York University Grossman School of
Medicine\\New York,\ 10016\\\asep\textbf{Erinn M. Hade}\\Department of
Population Health\\New York University Grossman School of Medicine\\New
York,\ 10016\\}
\date{2026-07-15}
\begin{document}
\maketitle
\begin{abstract}
Interrupted time series analysis has been used to model the effect of
policy and other interventions on public health by forecasting a
counterfactual time series during the intervention period using data
from prior to the intervention. However, due to typically relying on a
single study unit, this approach risks not adjusting for other
interruptions that precede and co-occur during the intervention period.
The COVID-19 pandemic is a prominent example of this phenomenon of
nuisance interruptions in contemporary public health research. To
address this complication, we propose using Bayesian stacking over a
range of functional forms for the impact of the nuisance interruption in
order to make counterfactual forecasts for the intervention period. We
used our proposed methods to estimate the impact of the 2021 Texas
six-week abortion ban on documented pregnancies among women in Texas
while adjusting for the impact of the COVID-19 pandemic.
\end{abstract}
{\bfseries \emph{Keywords}}
\def\sep{\textbullet\ }
Causal inference \sep Bayesian data analysis \sep Stacking \sep Model
combination \sep Electronic health records \sep 
Maternal health

\section{Introduction}\label{sec-intro}

The key feature of interrupted time series analysis (ITS) is the
occurrence of a single event or intervention that creates pre- and
post-interruption periods. A model that is fit using the
pre-interruption data can then be used to make a counterfactual forecast
without the interrupting event, allowing for inference of the event's
impact on outcomes. ITS analysis has been used to assess public health
interventions and policy changes
\citep{schafferInterruptedTimeSeries2021, papadogeorgou2023} and
pandemic impacts \citep{rileyImpactCOVID19Pandemic2022}, to perform
econometric analyses \citep{menchettiEstimatingCausalEffect2021} and A/B
testing \citep{brodersenInferringCausalImpact2015}.

While the research question of interest, ``What is the impact of the
interruption on outcomes?'' has been asked across many applications,
there is limited literature connecting ITS analysis to the conventional
potential outcomes framework
\citep{rubin1974, papadogeorgou2023, menchettiEstimatingCausalEffect2021}.
Moreover, there is a further gap in the literature for ITS methods with
a prior interruption (\emph{nuisance interruption}) to the event of
interest that is within the study time frame. These nuisance
interruptions have the potential to confound inferences in the primary
post-interruption period due to the interruptions' post-periods overlap.
restricting the study period to only analyze the post-nuisance
interruption data, this excludes information on the trend prior to the
nuisance event, limiting the statistical power and potential to pick up
on long-term trends such as seasonality. While a nuisance interruption
could also occur simultaneously with the intervention, the co-occurrence
poses an identifiability issue where we are unable to disentangle the
effects of the nuisance and intervention. In this paper, we limit our
focus to nuisance interruptions that occur strictly \emph{prior} to the
intervention.

Beyond potential for confounding by the nuisance interruption, it is
unclear what functional form the nuisance interruption's effect may
take. For example, consider investigating the impact of the 2021 Texas's
six-week abortion (Senate Bill8 or SB 8) on the rate of hypertensive
disorders of pregnancy (HDP). The ban was passed into law on July 1,
2021 and prohibits abortion after detection of fetal or embryonic
cardiac activity, generally around six weeks of gestation. The COVID-19
pandemic, which began in early 2020, has been observed to have a
negative impact on the rate of HDP
\citep{jacksonMaternalHealthCOVID192024}. If our goal is to estimate the
impact of Texas's six-week abortion ban on the rate of HDP over the time
period 2015 to 2024, the confounding interruption due to the pandemic
must be addressed in our model in order to estimate the ban's subsequent
impact.

To address potential confounding by a nuisance interruption with unknown
functional form, we propose a modeling strategy leveraging Bayesian
stacking over a range of time series models, each with a different
functional form of the nuisance interruption's effect
\citep{yaoUsingStackingAverage2018, hyndmanForecastingInterruptedTime2024}.
This approach is flexible in that it allows specification of various
time series models with valid predictive distributions and without
constraining the shape of the nuisance effect. Additionally, stacking
weights the nuisance effect models to optimize the out-of-sample
predictiveness on pre-intervention data making the best use of observed
data and expert knowledge.

The remainder of the paper is structured as follows: first, we provide
further background on the ITS methodology. Then, we discuss our proposed
methods for handling nuisance interruptions in the potential outcomes
framework. This includes discussion of the assumptions, what quantities
may be of interest, a particular specification for the outcome model,
and the details of Bayesian stacking. We then provide a simulation study
comparing the inferential abilities of the Bayesian stacking approach to
conventional ITS approaches such as Bayesian Structural Time Series and
a single covariate time series model. Next, we investigate the impact of
Texas's 2021 abortion ban on documented pregnancies using electronic
health record data from Epic Cosmos
\citep{tarabichiCosmosCollaborativeVendorFacilitated2021}. Finally, we
conclude with a discussion of the results and implications for ITS
analyses in related contexts.

\section{Background}\label{sec-background}

ITS analysis is useful for assessing policy impacts or natural
experiments
\citep{lopezbernalInterruptedTimeSeries2017, schafferInterruptedTimeSeries2021, jiangEstimatingEffectsHealth2022, papadogeorgou2023, bellTexas2021Ban2023, jacksonMaternalHealthCOVID192024}.
Previous approaches to ITS analysis have considered two related but
different strategies to make comparisons to a counterfactual prediction
\citep{jiangEstimatingEffectsHealth2022}. First, the fully parametric
prediction approach utilizes a time series model that is fit to the pre-
and post-intervention periods using some parametric functional form to
model the intervention impact in the post-intervention period
\citep{lopezbernalInterruptedTimeSeries2017}. The alternative model, the
semi-parametric forecasting approach, fits a time series model to the
pre-intervention data and then uses forecasting during the
post-intervention period to create a counterfactual time series
\citep{brodersenInferringCausalImpact2015, gianacasBayesianStructuralTime2023}.

The parametric prediction approach typically uses one of the following
models to estimate the intervention's effect on the outcome: segmented
(linear, logistic, Poisson) regression, ARIMA/ARMA modeling, generalized
additive (mixed) models (GAM), or Prais-Winsten regression models. Each
of these models parameterizes the effect as a step change, a slope or
ramp, as a pulse, or as some other readily defined functional form
\citep{lopezbernalInterruptedTimeSeries2017}. While not always
necessary, the counterfactual prediction can be made by setting the
intervention parameters to zero. Otherwise, the quantity of interest
could be the parameters themselves.

The semi-parametric forecasting approach fits a time series model to the
pre-intervention data and makes a counterfactual forecast for the
post-intervention period based solely on the pre-intervention data and
observed time-varying covariates assumed to be independent of the
intervention a particular functional form for the intervention's impact
\citep{brodersenInferringCausalImpact2015, gianacasBayesianStructuralTime2023}.
This approach assumes that trends pre-intervention would have continued
post-intervention in the counterfactual world where the intervention
does not occur. One particular model that has been used in this setting
is the Bayesian Structural Time Series (BSTS)
\citep{brodersenInferringCausalImpact2015}.

\begin{figure}

\centering{

\pandocbounded{\includegraphics[keepaspectratio]{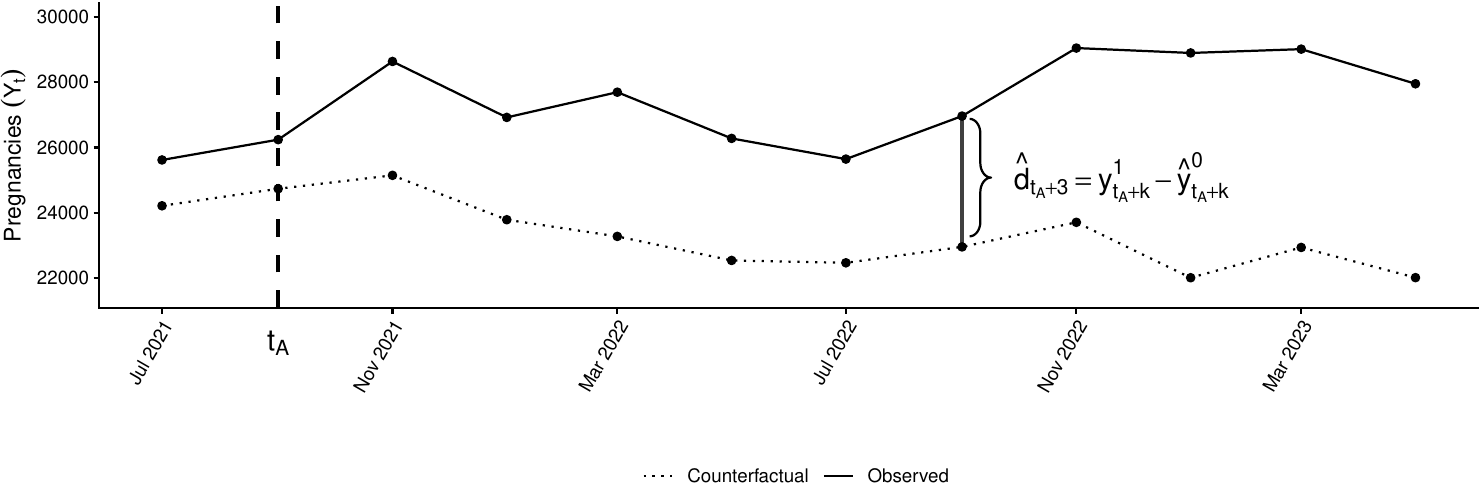}}

}

\caption{\label{fig-example-its}An illustration of how the point-wise
differences are calculated using the Texas documented pregnancies time
series and a counterfactual forecast from the semi-parametric negative
binomial model.}

\end{figure}%

As illustrated in Figure~\ref{fig-example-its}, the impact of the
intervention at any time \(t\) after the intervention begins at time
\(t_A\) is modeled as the difference between a counterfactual prediction
and observed data (\(\hat{d}_t = Y^1_t - \hat{Y}^0_t\)). The total
impact can be estimated as the cumulative sum of the point-wise
differences (\(\hat{D}_t = \sum_{t=t_A}^T \hat{d}_t\)) during the
post-intervention period. The estimands are defined in further detail in
the following section.

As noted by Gianacas et al.\citep{gianacasBayesianStructuralTime2023},
the fully parametric approach may require more effort to set up and
requires more assumptions about the shape of the intervention's impact,
but this approach can provide more stable estimates with smaller
standard errors than the semi-parametric approach. The semi-parametric
approach may be preferred when, (i) the intervention occurs later in the
time series rather than in the middle, (ii) there is potentially a lag
between implementation and impact, and (iii) there is a chance for model
misspecification\citep{jiangEstimatingEffectsHealth2022}. All three of
these apply to our application in the analysis of the impact of the 2021
Texas six week abortion ban on documented pregnancies among Texan women.

Further work formalized the causal assumptions and effects that are
incorporated in the proposed semi-parametric forecasting
approach\citep{papadogeorgou2023}. They term the setting as a single
unit time series where the unit represents a whole region. Their
proposal for the causal estimands, required causal assumptions, and
estimation methods form the basis for our framework as discussed in the
Methods section.

\section{Methods}\label{sec-methods}

We develop the causal framework and outcome model for both the fully
parametric and semi-parametric approaches to highlight their differences
and provide methods for when either approach is appropriate. The key
innovation we develop involves modeling the nuisance interruption that
occurs before the intervention and potentially continues during the
post-intervention period.

\subsection{Assumptions}\label{sec-methods-assume}

Papadoeorgou et al.\citep{papadogeorgou2023} propose two causal
identification assumptions that must be satisfied for estimates of a
causal effect on a single unit time series to be valid:

\begin{enumerate}
\def\labelenumi{\arabic{enumi}.}
\tightlist
\item
  The potential outcomes pre-intervention are unaffected by the
  intervention.
\item
  Covariates for predicting the outcome are unaffected by the
  intervention, and conditional on the covariates, the outcome series is
  stationary.
\end{enumerate}

The first assumption can always be satisfied by setting the intervention
initiation \(t_A\) to be a time earlier in the series such as when the
intervention was first proposed or passed. It can also be checked by
setting a fictional intervention time \(t_A^*\) prior to the true
initiation point \(t_A\) and estimating an effect at time \(t\) between
the fictional and true initiation point \(t \in [t_A^*, t_A].\) If a
null effect is consistent with the estimate, then the assumption is not
likely violated.

The second assumption allows for using pre-intervention data to forecast
a counterfactual in the post-intervention period. Violations of this
assumption can arise from two sources. First, if the covariates are
affected by the intervention, then the assumption is violated. This can
be checked by treating each covariate as an outcome and estimating the
intervention effect, where a null effect again indicates no violation.
The second source of violation comes from confounding in the series,
which results from systematic differences pre- and post-intervention
that are not due to the intervention itself \citep{papadogeorgou2023}.
In our analysis, the presence of the COVID-19 pandemic and its impact on
maternal health could act as a confounder that must be controlled for in
the model.

Further, we assume that the observed series \(Y_t\) during the
post-intervention period \(t \geq t_a\) is an observation of \(Y_t(1)\).

\subsection{Quantity of Interest}\label{quantity-of-interest}

Following the potential outcomes framework for ITS proposed
previously\citep{papadogeorgou2023}, for any given outcome \(Y\), the
quantity of interest will be the total count of events of \(Y\) due to
the intervention. Let \(Y_t(A_t)\) be the potential outcome count of
events of \(Y\) at time \(t\) with intervention indicator \(A_t\). Over
\(T\) time points, we observe \(Y_t(0)\) for \(t = 1,\dots, t_A-1\) and
\(Y_t(1)\) for \(t = t_A, \dots, T\). The unobserved counterfactual is
\(Y_t(0)\) for post-intervention period \(t \geq t_A\). Suppose we also
have fully observed covariates
\(\mathbf{X} \in \mathbb{R}^{T \times k}\) from both the pre- and
post-intervention periods. To emphasize that the quantity of interest is
not dependent on the model choice, we first develop the quantity we want
to estimate, before developing a model to perform the estimation.

To estimate the intervention effects via fully parametric modeling, the
data are not split into pre- and post-intervention segments; rather, the
model is fit to the whole data set. The intervention parameter estimates
are then directly used to estimate the intervention effect as the
difference between counterfactual means,

\begin{equation}\protect\phantomsection\label{eq-qoi-diff-means}{
\delta_t = E\left(Y_t(1) - Y_t(0) | \mathbf{y}_{1:(t_A-1)}, \mathbf{X}_{1:t}\right).
}\end{equation}

Semi-parametric estimation of the intervention effects begins with
splitting the data into pre- and post-intervention segments and fitting
the model only to the pre-intervention data. Next, we generate
post-intervention period forecasts for the counterfactual based on the
pre-intervention period model. The intervention effect estimate
\(\delta_t\) is non-parametric based on point-wise differences between
the predicted and observed counts in Equation~\ref{eq-qoi-diff-means}
and illustrated by Figure~\ref{fig-example-its}. The cumulative
intervention effect up to time \(t \geq t_A\) \(\Delta_t\) is,

\begin{equation}\protect\phantomsection\label{eq-cum-diff}{
\Delta_t = \sum_{i=t_A}^t E\left(Y_i(1) - Y_i(0) | \mathbf{y}_{1:(t_A-1)}, \mathbf{X}_{1:t}\right) = \sum_{i=t_A}^t \delta_i.
}\end{equation}

\subsection{An Outcome Model}\label{sec-modeling}

For the purposes of our analysis, we now propose a Bayesian
autoregressive Poisson outcome model. We adopt the Bayesian approach to
incorporate prior information, to implement Bayesian stacking, and to
take advantage of the probabilistic properties inherent to the posterior
predictive distribution.

The fully parametric model is an autoregressive Poisson generalized
linear model with with the log-link and an offset to model the outcome
as a rate, and step and slope changes for the intervention effect,

\begin{align*}
\log(\mu_t) = &\beta_0 + \mathbf{x}_t' \beta_1 + f_C(t, t_C, C_t; \beta) + \\
    & \beta_3 A_t + \beta_4 (t - t_A) A_t + \\
    & \sum_{p=1}^P \phi_p \log(Y_{t-p} + 1) + \log(W_t), \numberthis \label{eq-general-pois-ar-is}
\end{align*}

where

\begin{itemize}
\tightlist
\item
  \(f(t, t_C, C_t; \beta)\) is a function of the nuisance event's
  impact.
\item
  \(A_t\) is an indicator for the intervention.
\item
  \(t_A\) is the time period coinciding with the start of the
  intervention.
\item
  \(\mathbf{x}_t\) is a vector of time-varying predictors.
\item
  \(W_t\) is an offset to model the rates of the outcome.
\end{itemize}

The semi-parametric model drops the term
\(\beta_3 A_t + \beta_4 (t - t_A) A_t\) from the parametric model,

\begin{align*}
\log(\mu_t) = &\beta_0 + \mathbf{x}_t' \beta_1  + f(t, t_C, C_t; \beta) + \\
    & \sum_{p=1}^P \phi_p \log(Y_{t-p} + 1) + \log(W_t), \numberthis\label{eq-general-pois-ar-oos}
\end{align*}

The semi-parametric counterfactual forecast is independent of the
intervention conditional on the environment leading up to the
intervention. This assumes that those same conditions will continue
during the intervention period, controlling for the time-varying
covariates in \(\mathbf{x}_t.\)

Of note is that neither the quantity of interest nor the approach for
handling the nuisance interruption is dependent on the type of outcome
model, and an investigator could readily substitute a Gaussian, negative
binomial, zero-inflated Poisson, or various other types of outcome
models as long as they allow for including covariates, produce
forecasts, and have a valid predictive likelihood function. For example,
a negative binomial model with the GLM parameterization the {Equation
\ref{eq-nb-pmf}} could also use the same mean as in {Equations
\ref{eq-general-pois-ar-is} or \ref{eq-general-pois-ar-oos}} with an
additional overdispersion parameter \(r > 0\),

\begin{equation}
Pr(Y_t = y) = \frac{\Gamma(r + y)}{y!\Gamma(r)}\left(\frac{r}{r + \mu_t}\right)^r\left(\frac{\mu_t}{r + \mu_t}\right)^y, \label{eq-nb-pmf}
\end{equation}

\noindent for \(y = 0, 1, 2, \dots,\) such that
\(Var(Y_t) = \mu_t + \frac{\mu_t^2}{r}\).

Both the Poisson and negative binomial models use the same prior
distribution specifications for the model parameters
\(\theta = (\beta, \phi)'.\) The priors are jointly independent and
weakly informative; \(\beta_j \overset{iid}{\sim} N(0, 10^2)\) for
\(j = 0, \dots, 4\) and
\(\phi_{p} \overset{iid}{\sim} N_{[-1, 1]}(0, 1^2).\) In our
application, we investigated several potential prior distributions for
\(r\) such as the exponential, gamma, and Inverse-Gamma distributions in
a prior sensitivity analysis discussed in the appendix. Models are fit
using the Stan programming language and CmdStanr interface
\citep{standevelopmentteam2023, gabry2023}.

\subsection{Model Stacking over Nuisance Interruption Functional
Forms}\label{sec-nuisance}

In an analysis where a nuisance interruption occurs prior to the
intervention, model uncertainty is tied to the effect shape over time on
outcomes of both the nuisance interruption and the intervention
interruption. A range of effect shapes may be considered in a segmented
regression model, including a step change, slope change, both step and
slope, lagged changes, or temporary changes
\citep{lopezbernalInterruptedTimeSeries2017}. We can also adjust for the
nuisance interruption directly using similar adjustments. For example,
if the COVID-19 pandemic is a potential nuisance interruption, then we
can use the average COVID-19 case load or average number of deaths due
to COVID-19 over time as a direct measure of its potential impact.

Indeed, in our analysis of the effects of the six-week abortion ban on
documented pregnancies in Texas, we consider a range of effect
functional forms, which include the average COVID-19 case and death
loads for each bi-monthly period in our study from January 2020 to May
2023 \citep{dong2020, cdc2020}.

The nuisance interruption functional form \(f(.)\) can be manipulated
based on different assumptions of how the interruption could impact the
outcome. For our analysis, the direct measures of COVID-19 impact
include the presence of COVID-19 in the environment, time since the
beginning of the pandemic, current mean case load, current mean
percentage of deaths due to COVID-19, while functional forms include
step and/or slope changes, temporary step and/or slope changes, an
exponential decay, and a spike with a steady drop-off.

\begin{enumerate}
    \setlength{\itemsep}{0pt}
    \setlength{\parsep}{0pt}
    \item No Impact: $f(\mathbf{x_t}; \beta) = 0$
    \item Step: $f(\mathbf{x_t}; \beta) = \beta_1 C_t$
    \item Slope: $f(\mathbf{x_t}; \beta) = \beta_1 (t - t_C) C_t$ 
    \item Step and Slope: $f(\mathbf{x_t}; \beta) = \beta_1 C_t + \beta_2 (t - t_C) C_t$ 
    \item Step Ends: $f(\mathbf{x_t}; \beta) = \beta_1 C_t * (1 - A_t)$ 
    \item Slope Ends: $f(\mathbf{x_t}; \beta) = \beta_1 (t - t_C) C_t * (1 - A_t)$ 
    \item Step and Slope Ends: $f(\mathbf{x_t}; \beta) = \beta_1 C_t  * (1 - A_t) + \beta_2 (t - t_C) C_t * (1 - A_t)$ 
    \item Lagged Step: $f(\mathbf{x_t}; \beta) = \beta_1 C_{t-l}$ 
    \item Lagged Slope: $f(\mathbf{x_t}; \beta) = \beta_1 (t - t_C-l) C_{t-l}$ 
    \item Lagged Step and Slope: $f(\mathbf{x_t}; \beta) = \beta_1 C_{t-l} + \beta_2 (t - t_C - l) C_{t-l}$ 
    \item Pulse: $f(\mathbf{x_t}; \beta) = \beta_1 exp(-(t - t_C) / 2) C_t$ 
    \item Triangle: $f(\mathbf{x_t}; \beta) = \beta_1 (12 - |12 - t_C|)$ 
    \item Average Daily COVID-19 Cases: $f(\mathbf{x_t}; \beta) = \beta_1 M_t$, where $M_t$ is the daily mean cases of COVID-19 for each two-month period $t$ in Texas  
    \item Average Weekly \% of Deaths due to COVID-19: $f(\mathbf{x_t}; \beta) = \beta_1 D_t$, where $D_t$ is the mean of weekly \% deaths due to COVID-19 for each period $t$ in Texas. 
\end{enumerate}

We propose using Bayesian stacking to address model uncertainty while
appropriately incorporating uncertainty in the analysis estimates
\citep{yaoUsingStackingAverage2018}. Our innovation is to use stacking
with an ITS analysis to assess model uncertainty for the effect shape of
the intervention and uncertainty of the effect shape of the nuisance
interruption.

We perform stacking based on Pareto-Smoothed Importance-Sampling
Leave-Future-Out (PSIS-LFO) weights
\citep{yaoUsingStackingAverage2018, burknerApproximateLeavefutureoutCrossvalidation2020a, barigouBayesianModelAveraging2023}.
Previous work has developed the framework to calculate the
Leave-Future-Out cross validation (LFO-CV) values and the expected
log-predictive distribution (\(\mathrm{ELPD_{LFO}}\)) estimates from
log-likelihood values calculated at each step of a Markove Chain Monte
Carlo (MCMC) sampler \citep{bürkner2024}.

The log-likelihood of the data model is calculated at each observation
using the current posterior draws for each iteration of the sampler
generating \(S\) draws of the log-likelihood. From these, the PSIS-LFO
weights can be calculated which approximate the leave-future-out
posterior predictive distribution of each observation. These
approximations allow us to use the log-scoring rule (Kullback-Leibler
divergence) to find the optimal model weights to maximize the scoring
rule to compare the LFO posterior predictive distribution with the true
posterior predictive distribution. Equivalently, this minimizes the
divergence between the two predictive distributions.

To perform stacking, we first determine the model set \(\mathcal{M}\)
using the range of considered nuisance interruption effect shapes and
autoregressive (AR) orders \(p = 0,\dots, P\). For each
\(M_k \in \mathcal{M}\), we fit the model to the data to collect
posterior predictive distribution (PPD) draws
\(\mathbf{P}^k \in \mathbb{R}^{Q \times T}\) and log-likelihood values
\(\mathbf{L}^k = (\mathbf{l}_1, \dots, \mathbf{l}_T)'\) where
\(l_{i,q}^k = p_k(y_i | \theta^{(q)})\) and \(p_k(y | \theta)\) is the
log-likelihood for \(M_k\) evaluated at \(i = 1, \dots, T\) with
posterior draw \(\theta^{(q)}\) for \(q = 1,\dots, Q\). We use
approximate LFO-CV on the model fit to obtain the PSIS-LFO ELPD values
\(\mathrm{ELPD}_{LFO}\)
\citep{burknerApproximateLeavefutureoutCrossvalidation2020a, bürkner2024}.
Using the \(\mathrm{ELPD}_{LFO}\) values, model weights are calculated
using the stacking log scoring rule or Pseudo-Bayesian Model Averaging
(Pseudo-BMA)
\citep{yaoUsingStackingAverage2018, barigouBayesianModelAveraging2023}.

For the stacking rule, the weights \(\mathbf{w} = (w_1,\dots,w_K)'\) are
the solution to the optimization problem,

\begin{equation}\protect\phantomsection\label{eq-stacking-eqn}{
\mathbf{w} = \underset{w \in S_1^K}{\mathrm{argmax}} \sum_{i=L}^{N-M} \log \sum_{k=1}^K w_k p_k(y_{I+1:M} | y_{1:i}, M_k),
}\end{equation}

where
\(\mathcal{S}^K_1 = \{\mathbf{w} \in [0, 1]^K : \sum_{k=1}^K w_k = 1\},\)
\(N\) is the full series length, \(M\) is the number of steps ahead to
predict, \(L\) is the minimum series length to evaluate, and
\(p_k(y_{I + 1:M} | y_{1:i})\) is the predictive density for the
\(M\)-step ahead predictions given data up to time \(i.\)

For the pseudo-BMA rule, the weights are derived from the
\(\mathrm{ELPD}_{LFO}^k\) for each model \(k\),

\begin{equation}\protect\phantomsection\label{eq-pseudobma-eqn}{
w_k = \frac{\exp(\mathrm{ELPD}_{LFO}^k)}{\sum_{k=1}^K \exp(\mathrm{ELPD}_{LFO}^k)}.
}\end{equation}

The AR order is a primary part of the model that differentiates the
Poisson model used here from a conventional GLM with an offset. The AR
order is something that is typically selected using a model selection
technique like minimizing an information criterion (AIC, BIC, LOO-IC,
etc.) or by maximizing log-likelihood or \(R^2.\) The AR models of
different orders could also be included in the model stack by expanding
our range of considered models beyond various nuisance interruption
shapes.

\subsection{Cumulative Impact
Estimation}\label{cumulative-impact-estimation}

Suppose we have fit the Poisson semi-parametric model in {Equation
\ref{eq-general-pois-ar-oos}} to the pre-intervention data and obtained
posterior samples \(\tilde{\Theta} = (\tilde{\beta}, \tilde{\phi})\)
stored in a \(Q \times (K + 1 + P)\) matrix where \(Q\) is the number of
MCMC draws, \(K + 1\) is the number of \(\beta\) coefficients including
the intercept, and \(P\) is the AR order. Setting \(P = 1\), the \(q\)th
draw of parameters is
\(\tilde{\theta}^{(q)} = (\tilde{\beta}_0^{(q)}, \dots,\tilde{\beta}_K^{(q)}, \tilde{\phi}_1^{(q)})'\).
Then we obtain the posterior predictive draws given the pre-intervention
outcome and all covariates up to time \(t\),
\((\mathbf{y}_{1:(t_A-1)}, \mathbf{X}_{1:t}),\) as
\(\tilde{Y}_{t}^{(q)}(0)\) for \(t > T\) from
\(\tilde{Y}_{t}^{(q)}(0) | \mathbf{y}_{1:(t_A-1)}, \mathbf{X}_{1:t} \sim Pois(\tilde{\mu}_{t}^{(q)})\)
where

\begin{align*}
\log(\tilde{\mu}_{t}^{(q)}) = &\tilde{\beta}_0^{(q)} + \tilde{\beta}_1^{(q)_{'}} \mathbf{x}_t + f(t, t_C, C_t; \tilde{\beta}^{(q)}) + \\
    &\tilde{\phi}_1^{(q)} \log(\tilde{Y}_{t-1}^{(q)}(0) + 1) + \log(\mathrm{W_{t}}).\numberthis\label{eq-posterior-mean}
\end{align*}

Based on the previously mentioned causal assumptions, we can use the
posterior predictive forecast \(\tilde{Y}_{t}^{(q)}(0)\) for the
counterfactual series \(Y_{t}(0)\) for \(t \geq t_A\) to make inferences
on the intervention impacts, assuming that
\(\tilde{Y}_t(0)|\mathbf{y}_{1:(t_A-1)},\mathbf{X}_{1:t}\) is drawn from
the same distribution as
\(Y_t(0)|\mathbf{y}_{1:(t_A-1)},\mathbf{X}_{1:t}.\) Since the
counterfactual is approximated by posterior predictive MCMC draws, the
estimates derived from the counterfactual given the observed data
inherit the probabilistic properties of the posterior predictive
distribution.

The impact estimates are built upon the posterior predictive point
differences \(\tilde{d}_{t}^{q} = Y_{t} - \tilde{Y}_{t}^{(q)}(0)\) for
\(t \geq t_A.\) The differences are indexed by MCMC draw \(q\) and
stored in the \(Q \times (T - t_A)\) matrix
\(\tilde{\mathbf{D}} = (\tilde{\mathbf{d}}_{t_A}, \dots, \tilde{\mathbf{d}}_{T})\)
where
\(\tilde{\mathbf{d}}_{t} = \left(\tilde{d}_{t}^{(1)}, \dots, \tilde{d}_{t}^{(Q)}\right)'.\)

The mean posterior difference for \(\delta_t\) at time \(t\) is then
approximated by
\(\hat{\delta}_t = \frac{1}{Q} \sum_{q = 1}^Q \tilde{d}_{t}^{(q)}.\)
These differences can also be cumulatively summed to give the cumulative
impact up to time \(t\) as
\(\tilde{D}_t^{q} = \sum_{i = t_A}^t \tilde{d}_{i}^{(q)}\), which in
expectation estimates
\(\hat{\Delta}_{t} = \frac{1}{Q} \sum_{q = 1}^Q \tilde{D}_{t}^{(q)}.\)

Note that

\begin{align*}
E(\hat{\delta}_t | &\mathbf{y}_{1:(t_A-1)}, \mathbf{X}_{1:t}) = \frac{1}{Q}\sum_{q=1}^Q E\left(\tilde{d}_t^{(q)} | \mathbf{y}_{1:(t_A-1)}, \mathbf{X}_{1:t}\right) \\
&= E\left(Y_t - \tilde{Y}_t^{(q)}(0) | \mathbf{y}_{1:(t_A-1)}, \mathbf{X}_{1:t}\right) \\
&= E\left(Y_t(1) - Y_t(0) | \mathbf{y}_{1:(t_A-1)}, \mathbf{X}_{1:t}\right) \\
&= \delta_t, \numberthis \label{eq-exp-ptwise-diff}
\end{align*}

and

\begin{align*}
E(\hat{\Delta}_t | &\mathbf{y}_{1:(t_A-1)}, \mathbf{X}_{1:t}) = E\left(\sum_{i=t_A}^t \hat{\delta}_i | \mathbf{y}_{1:(t_A-1)}, \mathbf{X}_{1:t}\right) \\
&= \sum_{i=t_A}^t E\left(\hat{\delta}_i | \mathbf{y}_{1:(t_A-1)}, \mathbf{X}_{1:t}\right) \\
&= \sum_{i=t_A}^t \delta_i = \Delta_t. \numberthis \label{eq-exp-cumdiff}
\end{align*}

Similarly, we can estimate the posterior probability of an increase at a
given point \(t\geq t_A\) as
\(P(\delta_t > 0| \mathbf{y}_{1:(t_A-1)}, \mathbf{X}_{1:t}) = E(I(\delta_t > 0)| \mathbf{y}_{1:(t_A-1)}, \mathbf{X}_{1:t})\)
approximated by
\(\hat{p}_{t} = \frac{1}{Q} \sum_{q = 1}^Q I(\tilde{d}_{t}^{(q)} > 0).\)
Further, the estimated probability of an increase in the total
cumulative incidence of a particular maternal health outcome is
\(P(\Delta_t > 0| \mathbf{y}_{1:(t_A-1)}, \mathbf{X}_{1:t}) = E(I(\Delta_t > 0) | \mathbf{y}_{1:(t_A-1)}, \mathbf{X}_{1:t})\)
as
\(\hat{P}_{t} = \frac{1}{Q} \sum_{q=1}^Q I(\tilde{D}_{t}^{(q)} > 0).\)

\section{Simulation}\label{sec-simulation}

We performed a Monte Carlo simulation study to evaluate our proposed
method's inferential performance for estimating \(\Delta_T\) following
an intervention and a nuisance interruption prior to the intervention.
The study was designed to evaluate performance in settings similar to
the electronic health records (EHR) data analyzed in the application
case study; a short time series \(T = 39\), with a late interruption of
interest \(t_A = 30\), and a nuisance occurring 10 time steps prior to
the event \(t_C = 20\).

The data are generated using the Poisson AR(1) model given by {Equation
\ref{eq-general-pois-ar-is}} with parameters
\(\beta = (-2.2, 0.01, \beta_2)',\) \(\alpha,\) and \(\phi_1 = 0,\)
where \(\beta_2\) and \(\alpha\) are the confounding and intervention
strengths respectively. The confounding effect is either
\(f_1(C_t, t_C, \beta) = \beta_2 C_t\) or
\(f_2(C_t, t_C, \beta) = \beta_2 \left(0.2 \sin(\pi t / 4) + 0.3 \sin(\pi t / 8)\right).\)
The first effect \(f_1(C_t, t_C, \beta)\) is referred to as the
``simple'' confounding effect and is included in the set of averaged
models as well as being a submodel for the semi- and fully parametric
step-slope model. The second effect \(f_2(C_t, t_C, \beta)\) is referred
to as the ``complex'' confounding effect and is outside of the model
space used for stacking. Excluding the second effect from the model
space for the stacking methods and comparison models allows us to
evaluate how well the methods perform when they do not match the data
generating process.

We considered various confounding shapes (the ``simple'' step change vs
the ``complex'' sine wave), confounding strengths
(\(\beta_2 = 0, 0.01, 0.1\)), and intervention strengths
(\(\alpha = 0, 0.1\)). For brevity and readability, we present the
settings most similar to the Texas 2021 abortion ban analysis.

The proposed stacking Poisson ITS model was evaluated in both the
semi-parametric and fully parametric settings (abbreviated ``SP
Stacking'' and ``FP Stacking'' respectively) as discussed in the Methods
section. The model space included no nuisance interruption predictor
\(f_1\), a step predictor \(f_2\), step and slope predictors \(f_3\),
and a sinusoidal periodic predictor \(f_4\) following interruption,

\begin{align*}
f_1(t, t_C, C; \beta) &= 0 \\
f_2(t, t_C, C; \beta) &= \beta_1 C \\
f_3(t, t_C, C; \beta) &= \beta_1 C + \beta_3 t_C \\
f_4(t, t_C, C; \beta) &= \beta_1 \sin\left(\frac{\pi t_C}{4}\right),
\end{align*}

where \(t_C\) is the time of the nuisance event and
\(C_t= I(t \geq t_C)\) is an indicator for the nuisance event.

Five comparison methods were also evaluated, including a Bayesian
Structural Time Series model (BSTS) for ITS implemented by the
\texttt{CausalImpact} R package
\citep{brodersenInferringCausalImpact2015}, and semi- and fully
parametric versions of the Poisson ITS outcome model with either no
confounder covariate \(f_1\) (abbreviated ``SP No Conf'' and ``FP No
Conf'') or step-slope confounder covariates (``SP Step-Slope'' and ``FP
Step-Slope'') \(f_3\).

Inferential performance was evaluated using frequentist criteria
including the bias, 95\% credible interval coverage, and mean squared
error (MSE) of the cumulative difference estimate. Following the
guidelines for designing simulation studies\citep{morris2019}, we
include the estimated Monte Carlo uncertainty using the asymptotic
distributions for bias, 95\% confidence interval coverage, and MSE.
Formulas for the asymptotic confidence intervals are displayed in the
appendix. The simulation was run for 500 Monte Carlo samples for each
simulation setting.

\subsection{Simulation Results}\label{simulation-results}

\begin{figure}

\begin{minipage}[t]{\linewidth}

\centering{

\pandocbounded{\includegraphics[keepaspectratio]{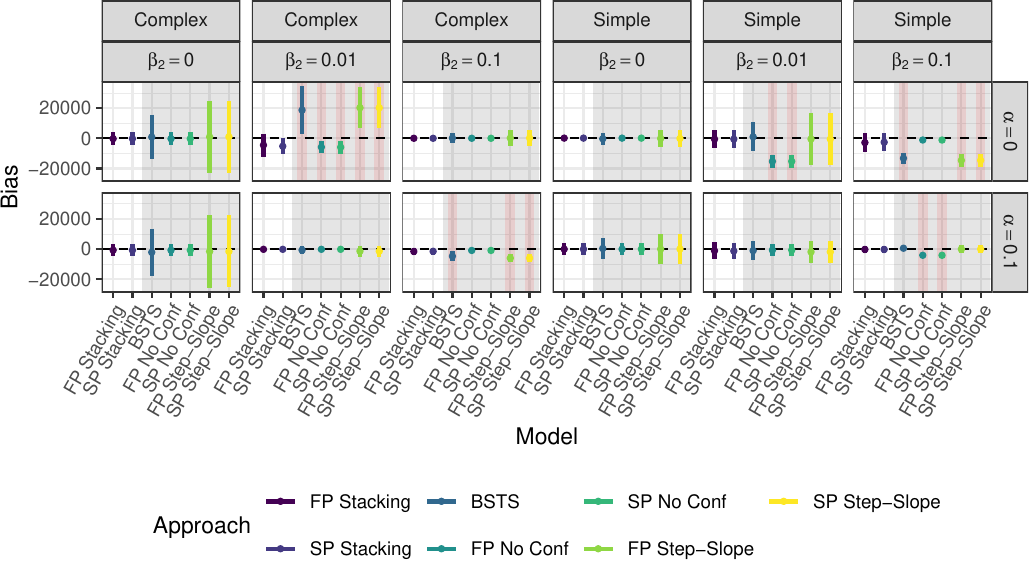}}

}

\subcaption{\label{fig-sim-bias-1}Bias in Cumulative Difference for
selected simulations.}

\end{minipage}%
\newline
\begin{minipage}[t]{\linewidth}

\centering{

\pandocbounded{\includegraphics[keepaspectratio]{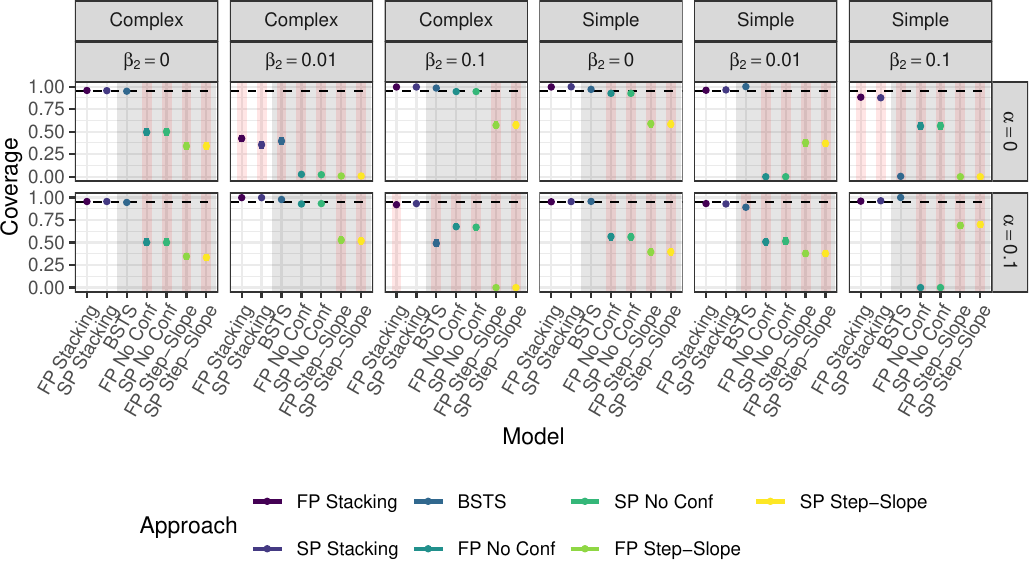}}

}

\subcaption{\label{fig-sim-bias-2}95\% Credible interval coverage of
Cumulative Difference for selected simulations.}

\end{minipage}%

\caption{\label{fig-sim-bias}Simulation study results.}

\end{figure}

Simulation results for the selected settings similar to the EHR data
(\(T = 39\), \(t_A = 30\), \(t_A - t_C = 10\)) are displayed as forest
plots in Figure~\ref{fig-sim-bias}. Estimates for each criterion (bias,
coverage, MSE, or credible interval width) are displayed with a point
for the mean and vertical bars covering the asymptotic 95\% confidence
interval. Stacking methods are highlighted with a white background while
comparison methods have a gray background. Methods whose 95\% confidence
intervals for bias do not include zero (horizontal dashed line) are
biased and are highlighted with a translucent vertical red bar.

{Table \ref{tab:sim-res-bias}} and Figure~\ref{fig-sim-bias-1} display
the estimated bias in estimating the cumulative difference for each
model under each setting. Across the selected settings, the SP Stacking
and FP Stacking methods perform well and are unbiased with relatively
narrow asymptotic confidence intervals for the bias. The BSTS and SP
Step-Slope and FP Step-Slope models perform similarly to each other in
most settings. These three methods are biased in several settings and
have very large asymptotic confidence intervals, far wider than either
stacking methods' intervals, indicating unstable point estimation in
general. The SP No Confounding and FP No Confounding models perform
similarly to each other, but are also biased in several settings, albeit
less so than the BSTS and SP Step-Slope and FP Step-Slope models. The No
Confounding models also have much narrower asymptotic confidence
intervals, similar to the two stacking methods, indicating more stable
point estimation.

The 95\% credible interval coverage of the true cumulative difference is
displayed in {Table \ref{tab:sim-res-ci}} and
Figure~\ref{fig-sim-bias-2}. A horizontal dashed line marks the nominal
95\% coverage level with translucent, vertical red bars highlighting
models whose 95\% confidence interval's upper bound for coverage does
not reach 95\%. For example, a coverage point estimate of 0.94 with 95\%
CI (0.925, 0.955) would be said to reach the nominal coverage despite
the point estimate being below 95\%.

Across the settings displayed in Figure~\ref{fig-sim-bias-2}, the SP and
FP Stacking methods reach nominal 95\% coverage except for the complex
confounding with a weak confounding effect and no intervention effect.
The BSTS method outperforms the other methods in some settings but
otherwise is similar to the Step-Slope methods, which fail to reach
nominal coverage for all of the presented settings. The No Confounding
methods are again similar and fail to reach nominal coverage for most
settings. The Step-Slope methods seem to fail to reach nominal coverage
due to high bias while the No Confounding methods seem to fail due to
too narrow credible intervals. The BSTS method reaches coverage where
Step-Slope methods do not, due to having too wide credible intervals to
overcome the large bias estimates noted in Figure~\ref{fig-sim-bias-1}.

\begin{table*}[htbp]

\caption{\label{tab:sim-res-bias}\label{tab:sim-res-bias}Mean bias for the cumulative differences and Monte Carlo standard errors for the bias for the cumulative differences in parentheses below the bias estimate. Biased estimates whose 95\% asymptotic confidence interval for bias does not contain zero are highlighted in bold.}
\centering
\begin{tabular}[t]{lccrrrrrrr}
\toprule
\multicolumn{4}{c}{ } & \multicolumn{2}{c}{Stacking} & \multicolumn{2}{c}{No Conf} & \multicolumn{2}{c}{Step-Slope} \\
\cmidrule(l{3pt}r{3pt}){5-6} \cmidrule(l{3pt}r{3pt}){7-8} \cmidrule(l{3pt}r{3pt}){9-10}
Conf. Shape & $\beta_2$ & $\alpha$ & BSTS & FP & SP & FP & SP & FP & SP\\
\midrule
Simple & 0 & 0 & -349 & 98.8 & 96.2 & 98.6 & 99.4 & -126.2 & -124.6\\
 &  &  & (2009.6) & (630.9) & (632.7) & (625.9) & (624.5) & (2852.5) & (2852.7)\\
Simple & 0 & 0.1 & 418.7 & -6.2 & -18.2 & -12 & -11.9 & 4.2 & 75\\
 &  &  & (3583.1) & (2143.6) & (2114.6) & (2070.5) & (2074.9) & (5165.4) & (4982.1)\\
Simple & 0.01 & 0 & 971.6 & -566.6 & -564.8 & \textbf{-15332.3} & \textbf{-15330} & -501.8 & -566.6\\
 &  &  & (4888.8) & (3036.6) & (3001.8) & (2319.8) & (2305.5) & (8890.8) & (8849.6)\\
Simple & 0.01 & 0.1 & -1025.9 & -1074.8 & -1234 & -670.2 & -664.7 & -1834 & -1832.4\\
 &  &  & (3112.3) & (2947.7) & (2838.2) & (2084.4) & (2078.1) & (3820.9) & (3825.4)\\
Simple & 0.1 & 0 & \textbf{-13258.2} & -2872.4 & -2619.6 & -1168.7 & -1176.4 & \textbf{-14661.5} & \textbf{-14668}\\
 &  &  & (1926) & (3322.5) & (3086.4) & (696.9) & (697.1) & (2136.3) & (2131.9)\\
Simple & 0.1 & 0.1 & 600.1 & -214.1 & -196.6 & \textbf{-4080.7} & \textbf{-4078.4} & 7.6 & 10.6\\
 &  &  & (781.9) & (1154.8) & (1131.5) & (701.5) & (701) & (1424.5) & (1422.2)\\
Complex & 0 & 0 & 803.7 & -290.1 & -283.7 & -284.5 & -292 & 903.7 & 945.3\\
 &  &  & (7583.2) & (2155.1) & (2167.1) & (2156.7) & (2159.9) & (12287.4) & (12288.2)\\
Complex & 0 & 0.1 & -2113.5 & -687.9 & -693.5 & -707.1 & -708.6 & -1586.4 & -1501.1\\
 &  &  & (8021.8) & (2113.9) & (2100.8) & (2083.1) & (2084.2) & (12251.1) & (12212)\\
Complex & 0.01 & 0 & \textbf{18684.9} & -4535.7 & -5270.8 & \textbf{-5873.3} & \textbf{-5961} & \textbf{20244.1} & \textbf{20202.8}\\
 &  &  & (8009.1) & (3914.2) & (2713.1) & (2108.9) & (2119.8) & (6988.4) & (6998.1)\\
Complex & 0.01 & 0.1 & -666.2 & -130.8 & -132.8 & -125.5 & -124.3 & -1343.1 & -1343.6\\
 &  &  & (1376.2) & (696.6) & (688) & (677.7) & (680) & (1874.5) & (1876.3)\\
Complex & 0.1 & 0 & 124.8 & -6.3 & -10.5 & -9.4 & -9.9 & 135.6 & 140.3\\
 &  &  & (1578.3) & (630.1) & (625.7) & (627.1) & (623.8) & (2619.3) & (2631.4)\\
Complex & 0.1 & 0.1 & \textbf{-4679.6} & -1544.2 & -1458.9 & -864.5 & -870.2 & \textbf{-5989} & \textbf{-5992.9}\\
 &  &  & (1646.9) & (1337) & (1285) & (664.5) & (664.9) & (1380.9) & (1375.8)\\
\bottomrule
\end{tabular}
\end{table*}

\begin{table*}[htbp]

\caption{\label{tab:sim-res-ci}\label{tab:sim-res-ci}Mean 95\% credible interval coverage for the cumulative differences and with Monte Carlo standard errors in parentheses below each coverage estimate.Models and settings with nominal 95\% or higher coverage are highlighted in bold.}
\centering
\begin{tabular}[t]{lccrrrrrrr}
\toprule
\multicolumn{4}{c}{ } & \multicolumn{2}{c}{Stacking} & \multicolumn{2}{c}{No Conf} & \multicolumn{2}{c}{Step-Slope} \\
\cmidrule(l{3pt}r{3pt}){5-6} \cmidrule(l{3pt}r{3pt}){7-8} \cmidrule(l{3pt}r{3pt}){9-10}
Conf. Shape & $\beta_2$ & $\alpha$ & BSTS & FP & SP & FP & SP & FP & SP\\
\midrule
Simple & 0 & 0 & \textbf{0.97} & \textbf{0.996} & \textbf{0.998} & 0.925 & 0.925 & 0.588 & 0.582\\
 &  &  & (0.008) & (0.003) & (0.002) & (0.012) & (0.012) & (0.022) & (0.022)\\
Simple & 0 & 0.1 & \textbf{0.957} & \textbf{0.952} & \textbf{0.954} & 0.566 & 0.562 & 0.395 & 0.397\\
 &  &  & (0.009) & (0.01) & (0.01) & (0.023) & (0.023) & (0.023) & (0.023)\\
Simple & 0.01 & 0 & \textbf{1} & \textbf{0.96} & \textbf{0.964} & 0 & 0 & 0.378 & 0.37\\
 &  &  & (0) & (0.009) & (0.008) & (0) & (0) & (0.022) & (0.022)\\
Simple & 0.01 & 0.1 & 0.891 & \textbf{0.932} & \textbf{0.928} & 0.508 & 0.514 & 0.377 & 0.377\\
 &  &  & (0.014) & (0.011) & (0.012) & (0.023) & (0.023) & (0.022) & (0.022)\\
Simple & 0.1 & 0 & 0.004 & 0.883 & 0.877 & 0.561 & 0.565 & 0 & 0\\
 &  &  & (0.003) & (0.014) & (0.015) & (0.022) & (0.022) & (0) & (0)\\
Simple & 0.1 & 0.1 & \textbf{1} & \textbf{0.96} & \textbf{0.964} & 0 & 0 & 0.688 & 0.7\\
 &  &  & (0) & (0.009) & (0.008) & (0) & (0) & (0.021) & (0.021)\\
Complex & 0 & 0 & \textbf{0.949} & \textbf{0.957} & \textbf{0.955} & 0.494 & 0.5 & 0.34 & 0.34\\
 &  &  & (0.01) & (0.009) & (0.009) & (0.022) & (0.022) & (0.021) & (0.021)\\
Complex & 0 & 0.1 & \textbf{0.945} & \textbf{0.955} & \textbf{0.955} & 0.501 & 0.501 & 0.347 & 0.335\\
 &  &  & (0.01) & (0.009) & (0.009) & (0.023) & (0.023) & (0.021) & (0.021)\\
Complex & 0.01 & 0 & 0.396 & 0.424 & 0.355 & 0.026 & 0.022 & 0.008 & 0.008\\
 &  &  & (0.022) & (0.022) & (0.022) & (0.007) & (0.007) & (0.004) & (0.004)\\
Complex & 0.01 & 0.1 & \textbf{0.977} & \textbf{0.998} & \textbf{0.998} & \textbf{0.928} & \textbf{0.932} & 0.528 & 0.52\\
 &  &  & (0.007) & (0.002) & (0.002) & (0.012) & (0.011) & (0.023) & (0.023)\\
Complex & 0.1 & 0 & \textbf{0.984} & \textbf{0.996} & \textbf{0.996} & \textbf{0.945} & \textbf{0.945} & 0.57 & 0.574\\
 &  &  & (0.006) & (0.003) & (0.003) & (0.01) & (0.01) & (0.022) & (0.022)\\
Complex & 0.1 & 0.1 & 0.49 & 0.92 & \textbf{0.932} & 0.676 & 0.67 & 0 & 0\\
 &  &  & (0.022) & (0.012) & (0.011) & (0.021) & (0.021) & (0) & (0)\\
\bottomrule
\end{tabular}
\end{table*}

\section{Impacts of the 2021 Texas Six-week Abortion Ban on
Pregnancies}\label{sec-example}

We now demonstrate our proposed methods using data from Texas, which
passed a six-week abortion ban through SB8 in July 2021. These data come
from Epic's Cosmos database, which harmonizes electronic health records
from Epic EHR systems across hospitals and clinics into a deidentified
data source \citep{tarabichiCosmosCollaborativeVendorFacilitated2021}.

\subsection{Data}\label{data}

While Cosmos data are well curated and available to researchers at
participating institutions, we note several characteristics to consider
when using these data. First, patient state of residence is the most
recent address recorded in Cosmos relative to the date of data
extraction. Our analyses include women whose address in the EHR was in
Texas as of December 29, 2024. \emph{Within} study timeline (January
2017 to June 2023), these women may have previously received care and
lived in states other than Texas. Additionally, the composition of the
patient population contributing data to Cosmos varies over time as
health systems/clinics are onboarded. As clinical site is not available
in these de-identified data, it is not possible to use clinic-related
data to adjust for the shifting demographics over time.

Finally, due to data privacy required by Cosmos for aggregated data,
small cell counts are masked. For any given subset of data used for
analysis (in our case,
\(\mbox{time frame} \times \mbox{location} \times \mbox{outcome} \times \mbox{demographics/insurance}\)),
the count of occurrences \(X\) is only reported if \(X > 10\) or if
\(X = 0\). This leads to considerable censoring, e.g.~for low population
states, rare events, small socio-demographic groups, or rare insurance
types (e.g., self-pay). Due to these reporting requirements, to increase
counts, we requested Cosmos data aggregated into six, two-month groups
per year from January-February 2017 to May-June 2023, resulting in
\(T = 39\) time points.

The demographic changes noted across the study period in the Texas
Cosmos population roughly match the U.S. Census Bureau's American
Community Survey (ACS) for age groups and race and ethnicity groups
(assuming two or more races is similar to Hispanic in ACS's racial
categories). Assessment of insurance over time between these data
sources is challenging since the ACS reports insured/not insured without
more detail of insurance type. The percentage of insured remained steady
from 2017 to 2023 in the ACS data, while the proportion of insured in
Cosmos varied over time in Texas.

The COVID-19 positive case counts and death data were collected from the
Johns Hopkins CSSE COVID-19 Data Repository and Center for Disease
Control \citep{dong2020, cdc2020}.\footnote{Cases data from
  \href{https://github.com/CSSEGISandData/COVID-19/blob/master/csse_covid_19_data/csse_covid_19_time_series/time_series_covid19_confirmed_US.csv}{Johns
  Hopkins CSSE} and deaths data from
  \href{https://covid.cdc.gov/covid-data-tracker/\#trends_weeklydeaths_weeklypctdeaths_35}{CDC}.}

As a case study and demonstration of our proposed methods, we present
the results for the impact of the 2021 six-week abortion ban on
documented pregnancies for women of reproductive age in Texas. This
outcome was chosen from among other reproductive health outcomes to
highlight the modeling choices that must be made regarding the space of
nuisance functionals, time-varying covariates, and overdispersion.
However, our outcome measure still has significant caveats. The
pregnancy rate per 1,000 women is based on the occurrence of pregnancy
diagnostic codes and the count of women of reproductive age in the
Cosmos sites in Texas for each two-month period. The pregnancy frequency
is a proxy for the number of pregnancies beginning in each two-month
period in Texas for women of reproductive age who received care at a
medical facility whose records connect to Epic Cosmos.

\subsection{Application Models}\label{application-models}

We performed the above-described analysis using Poisson and negative
binomial models for the fully parametric and semi-parametric approaches,
resulting in four analyses.

\begin{table*}

\caption{\label{tbl-trend-pred}The set of socio-demographic covariates and time $t$, and seasonality $t_S = \sin\left(\pi(t-2017)/3 + \pi/2\right)$ trend predictors included in the model space.}

\centering{

\begin{tabular}{ll}
\hline
 Name & $\mathbf{x}_t'\beta$ \\
\hline
Time Overall & $\beta_1 t$ \\
Time Seasonal & $\beta_1 t_S$ \\
Time Overall \& Seasonal & $\beta_1 t + \beta_2 t_S$ \\
Race \& Ethnicity; Time Overall \& Seasonal & $\beta_1 t + \beta_2 t_S + \beta_3 R_{t,W} + \beta_4 R_{t,B} + \beta_5 R_{t,H}$ \\
Age; Time Overall \& Seasonal & $\beta_1 t + \beta_2 t_S + \beta_3 A_t$ \\
Insurance type; Time Overall \& Seasonal & $\beta_1 t + \beta_2 t_S + \beta_3 I_{t,P} + \beta_4 I_{t,M}$ \\
All & $\beta_1 t + \beta_2 t_S + \beta_3 R_{t,W} + \beta_4 R_{t,B} + \beta_5 R_{t,H} +$ \\
&  $~~~~\beta_6 A_t + \beta_7 I_{t,P} + \beta_8 I_{t,M}$ \\
\hline
\end{tabular}

}

\end{table*}%

We based our model space for the model averaging on the Cartesian
product of the set of COVID-19 impact functions \(A = \{\) No Impact,
Step,\ldots, Avg. Weekly Deaths \(\}\) listed previously and a set of
time-varying socio-demographic covariates, and seasonality and trend
predictors \(B = \{\) Time, Seasonal,\ldots, All \(\}\) displayed in
Table~\ref{tbl-trend-pred}. The socio-demographic predictors include
\(R_{t, W}\), \(R_{t, B}\), and \(R_{t,H}\) which are the proportions of
Texas women in Cosmos during period \(t\) who are White, Black, or
Hispanic respectively; \(A_t\) which is the proportion of Texas women in
Cosmos who are 35 years of age or older; and \(I_{t,P}\) and \(I_{t,M}\)
which are the proportion of Texas women in Cosmos with public or private
insurance in period \(t\) respectively. Each model in the model space is
a concatenated pair from
\(M \in \mathcal{M} = \mathrm{concat}(A \times B) = \{\) Time+No Impact,
Seasonal+No Impact,\ldots, All+Avg. Weekly Deaths \(\}\) where the
\(\mathrm{concat}()\) function additively joins the two elements of each
ordered pair in a set. In total, the 14 COVID-19 pandemic impact
functions and 7 predictor functions created a set of 98 sub-models that
were then weighted by stacking.

The analysis was performed twice for each model family, once using the
semi-parametric approach, and again using the fully parametric approach
in order to demonstrate the potential for different results depending on
the approach. The results of the analyses are displayed in
{Figure~\ref{fig-overlay-all-tx} and Table \ref{tab:cumdiff-tx}}. The
LFO-IC and model stacking model weights for each sub-model are available
as CSV files in the supplementary materials.

The Poisson and negative binomial models use jointly independent, weakly
informative priors for \(\beta_j \overset{iid}{\sim} N(0, 10^2)\) for
\(j = 0, \dots, 4\) and
\(\phi_{p} \overset{iid}{\sim} N_{[-1, 1]}(0, 1^2).\) The negative
binomial model also requires specifying an additional prior for the
overdisperson parameter \(r\). To choose this prior, we performed a
sensitivity analysis comparing the posterior distribution of \(r\) under
exponential and Inverse-Gamma distribution priors with different fixed
scale parameter values \(\theta = 0.1, 1, 10, 100, 1000\) and with a
hyper-prior on the scale parameter, \(\theta \sim Inv-Gamma(3, 2\eta)\)
where \(\eta = E(\theta) = 0.1, 1, 10, 100, 1000\).

In our analysis we used the Inverse-Gamma prior with a hyper-prior on
the scale parameter where the scale parameter has prior expectation
\(E(\theta) = 10\). See the appendix for further details on the
sensitivity analysis.

\subsection{Application Results}\label{application-results}

\begin{figure}[htbp]

\begin{minipage}[t]{0.48\linewidth}

\centering{

\pandocbounded{\includegraphics[keepaspectratio]{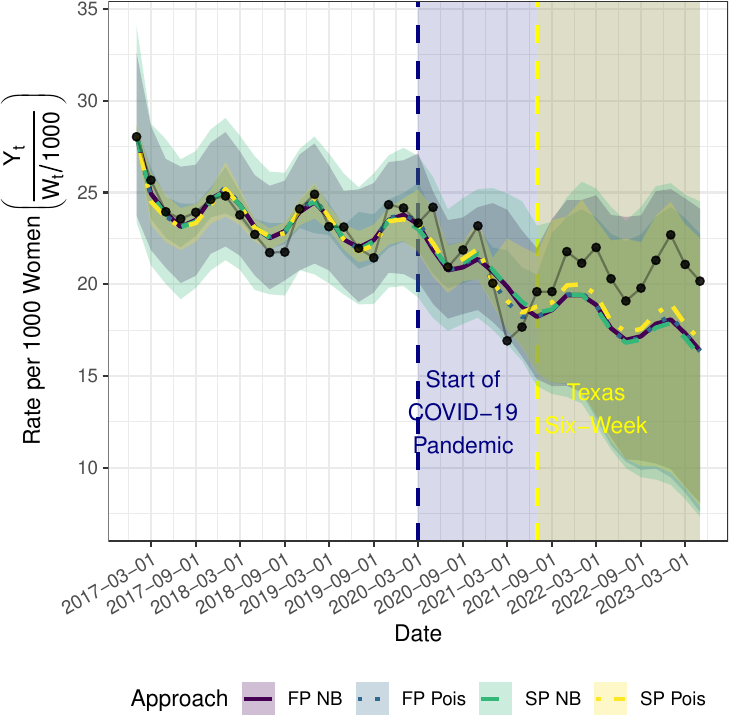}}

}

\subcaption{\label{fig-overlay-all-tx-1}Documented pregnancy rates per
1000 pregnancies.}

\end{minipage}%
\begin{minipage}[t]{0.04\linewidth}
~\end{minipage}%
\begin{minipage}[t]{0.48\linewidth}

\centering{

\pandocbounded{\includegraphics[keepaspectratio]{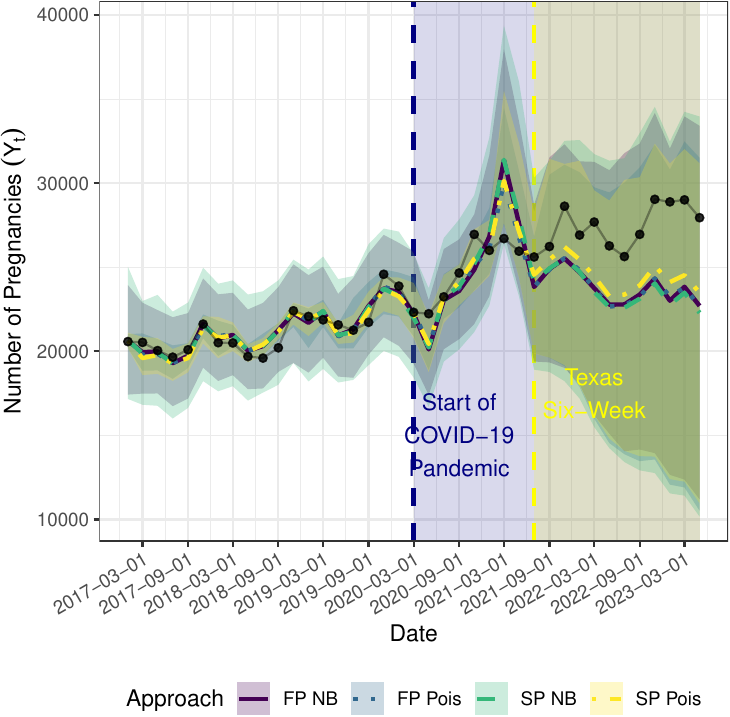}}

}

\subcaption{\label{fig-overlay-all-tx-2}Documented pregnancy counts.}

\end{minipage}%
\newline
\begin{minipage}[t]{0.48\linewidth}

\centering{

\pandocbounded{\includegraphics[keepaspectratio]{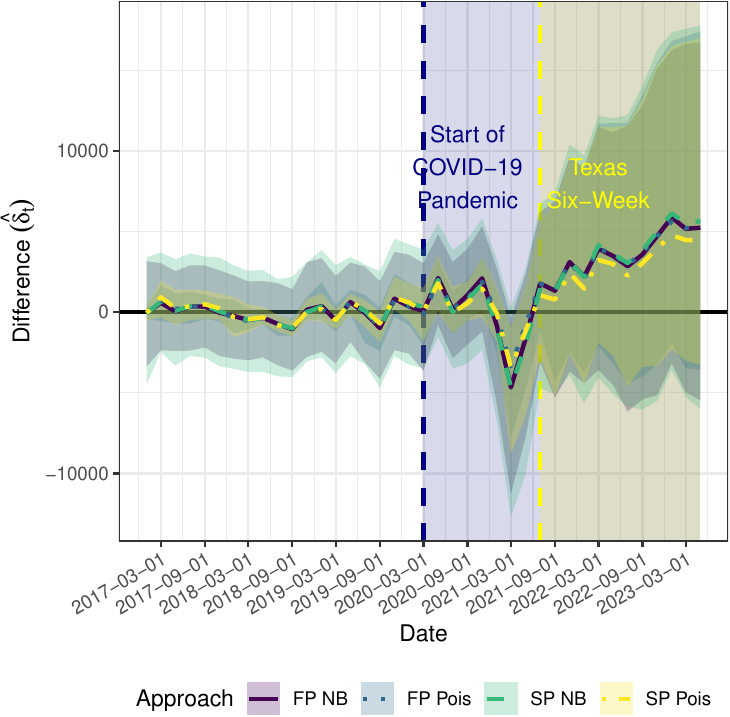}}

}

\subcaption{\label{fig-overlay-all-tx-3}Differenced counts.}

\end{minipage}%
\begin{minipage}[t]{0.04\linewidth}
~\end{minipage}%
\begin{minipage}[t]{0.48\linewidth}

\centering{

\pandocbounded{\includegraphics[keepaspectratio]{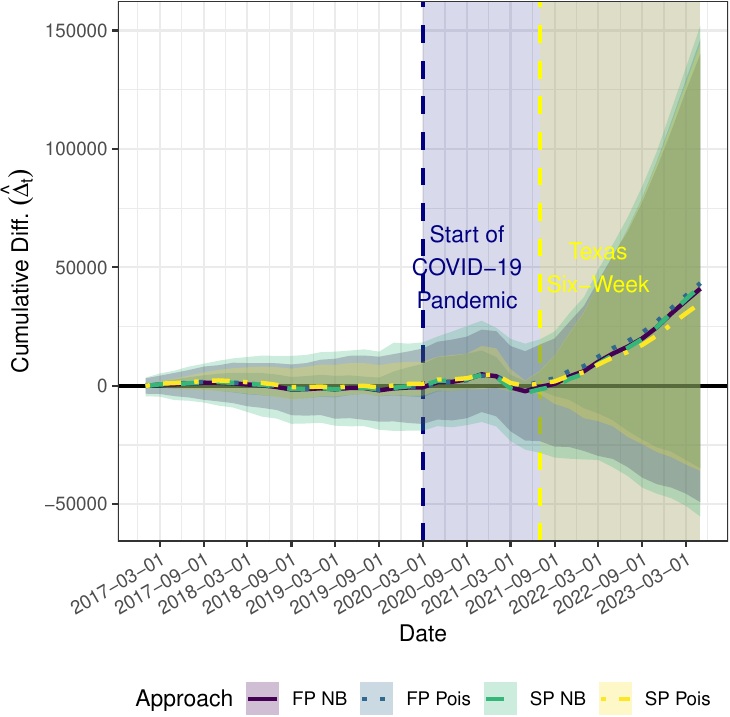}}

}

\subcaption{\label{fig-overlay-all-tx-4}Cumulative differences.}

\end{minipage}%

\caption{\label{fig-overlay-all-tx}Overlaid rates, counts, differences,
and cumulative differences in pregnancies in Texas with FP Stacking and
SP Stacking for the Poisson and negative binomial models. The gray line
and points in (a) and (b) are the observed data.}

\end{figure}%

\begin{table*}[htbp]

\caption{\label{tab:cumdiff-tx}\label{tab:cumdiff-tx}Cumulative differences between the documented pregnancy counts in Texas and the counterfactuals.}
\centering
\begin{tabular}[t]{lrrrrr}
\toprule
Model & Estimate & SD & 95\% CI LB & 95\% CI UB & $P(CumDiff>0)$\\
\midrule
Semi-Parametric Poisson & 34958 & 48084 & -35877 & 144316 & 0.76\\
Fully Parametric Poisson & 43326 & 42404 & -34857 & 146528 & 0.90\\
Semi-parametric NB & 42082 & 56332 & -55232 & 152098 & 0.77\\
Fully parametric NB & 41032 & 50109 & -49163 & 140182 & 0.82\\
\bottomrule
\end{tabular}
\end{table*}

In {Figure~\ref{fig-overlay-all-tx}@}, a dark blue vertical line
indicates the beginning of the COVID-19 pandemic, corresponding to the
first bi-monthly period (March-April 2020) where pregnancies could have
been impacted by the pandemic. Similarly, a yellow vertical line marks
the first bi-monthly period for conception of pregnancies potentially
impacted by the Texas six-week abortion ban (July-August 2021). The
periods after each line are shaded to indicate these times are
potentially impacted by each event.

Our posterior mean estimates for the number of excess pregnancies during
the ban period range from 34,958 excess pregnancies with the
semi-Parametric Poisson model to 43,326 excess pregnancies with the
fully parametric Poisson model. With all four modeling approaches, the
95\% credible intervals for the posterior number of excess pregnancies
are extremely wide by the end of the study period (roughly 180,000 to
210,000 pregnancies). This is due to the accumulated uncertainty via
summing \(\hat{\delta}_t\).

We also found there to be between a 76\% and 90\% probability of an
increase in excess pregnancies during the ban period.

\section{Discussion}\label{sec-discussion}

Nuisance interruptions to time series complicate ITS analysis of
intervention effects by potentially confounding the intervention of
interest through time. The co-occurrence of the interruptions calls for
adjustment for the confounder's effect. We proposed to address the
nuisance interruption confounding using stacking over a model space of
the potential functional forms for the nuisance interruption's effect,
and to estimate the intervention's effect using posterior estimates for
the point-wise or cumulative difference between the observed time series
and a forecast or predicted counterfactual.

In our simulations, we found that stacking in either the semi-parametric
or fully parametric model performed well in terms of bias and had
reasonable coverage across scenarios. Given the flexibility of stacking
models to accommodate varying shapes of the nuisance interruption, we
found these methods had least bias and reached nominal coverage across
simulation scenarios, outperforming comparison methods.

Despite similarities between FP and SP Stacking in our simulation
results, credible interval coverage, and credible interval widths for
the estimation of the cumulative difference, we have seen more
substantial differences between the two methods in practice. As such,
the choice between the SP and FP Stacking methods should be motivated by
the level of confidence in choosing a functional form for the
intervention effect. If the functional form is fairly certain, then the
FP Stacking method is suitable, otherwise SP Stacking should be used.
Graphical checks for the quality of the prediction or forecast should
also be performed to ensure that the counterfactual is plausible.

The Bayesian structural time series (BSTS) method has become a standard
ITS method in recent years. However, in our simulation study, BSTS
performed poorly with regard to bias, credible interval width, and MSE
in estimating the cumulative difference due to the intervention. While
this may be partly due to model mis-specification (the
\texttt{CausalImpact} package only allows for normally distributed
outcomes), we would expect this method to underestimate the variance
that was seen in the simulations. We would advise caution when using
this approach to ITS analysis when there is a confounding interruption.

In our semi-parametric negative binomial analysis of excess observed
pregnancies in Texas following the 2021 abortion ban, we found weak
evidence of an increase in observed pregnancies for women of
reproductive age with a cumulative estimate of 42,082 excess pregnancies
and a 95\% credible interval of (-55,232, 152,098) pregnancies. There
was a corresponding estimated 77\% posterior probability of a cumulative
increase in pregnancies following the ban. For comparison, the fully
parametric Poisson model estimated a similar increase of 43,326
pregnancies, a 95\% CI of (-34,857, 146,528) pregnancies, and a 90\%
posterior probability of a cumulative increase.

Recent work describing challenges in the evaluation of abortion
restrictions on outcomes highlights the COVID-era disruptions and the
need to account for the impact of COVID-19 and the recovery from the
pandemic when estimating the abortion restriction effect
\citep{gemmillMethodologicalConsiderationsInvestigating2026}. Analyses
measuring the impact of the 2021 Texas abortion ban on infant mortality
\citep{gemmillInfantDeathsTexas2024, bellTexas2021Ban2023, gemmillUSAbortionBans2025}
and fertility \citep{bellUSAbortionBans2025} did not directly account
for COVID-19 pandemic effects on their respective outcomes. Instead they
attempted to account for COVID-19 effects through comparison to other
states, or side-stepped the issue by performing a cross-sectional
analysis
\citep{stevensonTrendsMaternalDeath2024, thornburgAnxietyDepressionSymptoms2024}.
ITS analysis was previously used to investigate impact of the COVID-19
pandemic itself on maternal health, demonstrating the method's utility
in estimating public health impacts following disruptive events or
interventions \citep{jacksonMaternalHealthCOVID192024}. Our proposed
method provides an approach to directly account for the COVID-19
pandemic in assessments of abortion ban impacts, as well as other
instances when nuisance interruptions precede the intervention of
interest.

\subsubsection*{Acknowledgments}
    The computational work performed on this project was done with help
    from the NYU Big Purple High-Performance Computing cluster.

\subsubsection*{Conflicts of Interest Disclosure}
    We have no conflicts of interest to disclose.

\subsubsection*{Funding}
This work was funded by the Educational Foundation of America (EFA).

This research is also supported in part by an NYU CTSA grant UM1TR005769
from the National Center for Advancing Translational Sciences, National
Institutes of Health.

\subsubsection*{Supplementary Materials}
The LFO-IC and model weight tables for the Poisson and negative binomial
outcome models from the pregnancy analysis are available as .CSV files.
R code for the simulations and ITS analysis are available upon request.

\renewcommand\refname{References}
\bibliography{ref.bib}

\newpage\clearpage
\appendix
\counterwithin*{equation}{section}
\renewcommand{\theequation}{A\arabic{equation}}
\renewcommand{\thefigure}{A\arabic{figure}}
\renewcommand{\thetable}{A\arabic{table}}
\setcounter{equation}{0}
\setcounter{figure}{0}
\setcounter{table}{0}
\section{Appendix}
\subsection{Asymptotic Confidence Intervals for Simulation Evaluation
Criteria}\label{sec-app-eval-ci}

Let the true cumulative difference be \(\Delta_T\) with posterior mean
estimate \(\hat{\Delta}_{T,i}^{m}\) for estimation method \(m \in \{\)
BSTS, SP No Conf, FP No Conf, SP Step-Slope, FP Step-Slope, SP Stacking,
FP Stacking\(\}\) in the \(i\)-th simulation iteration. Also, let
\(\hat{\Delta}_{T,q, i}^{(m)}\) be the \(q\)-th quantile of the
posterior for \(\Delta_T\) with method \(m\) in the \(i\)-th simulation
iteration. For each criterion, we assume asymptotic normality based on
the central limit theorem to construct typical Wald confidence intervals
for each criteria\citep{morris2019}.

We reproduce the formulas from Morris et al.\citep{morris2019} for bias
and confidence interval coverage here. The estimate for bias is
\(\hat{\mathrm{Bias}} = \frac{1}{n_{sim}} \sum_{I=1}^{n_{sim}} \hat{\Delta}_{T,i}^{(m)} - \Delta_T\)
with Monte Carlo standard error
\(SE(\hat{\mathrm{Bias}}) = \sqrt{\frac{1}{n_{sim}(n_{sim}-1)} \sum_{I=1}^{n_{sim}} (\hat{\Delta}_{T,i}^{(m)} - \Delta_T)^2}\).
For 95\% CI coverage, the point estimate is
\(\hat{\mathrm{Cov}} = \frac{1}{n_{sim}} \sum_{I=1}^{n_{sim}} I(\hat{\Delta}_{T,0.025, i}^{(m)} \leq \Delta_T \leq \hat{\Delta}_{T, 0.975, i}^{(m)})\)
with standard error
\(SE(\hat{\mathrm{Cov}}) = \sqrt{\frac{1}{n_{sim}} \hat{Cov}^m (1 - \hat{Cov}^m)}\).

\subsection{Prior Sensitivity Analysis}\label{sec-app-priors}

We performed a prior sensitivity analysis for the negative binomial
autoregressive model to assess the sensitivity to the prior for the
overdispersion parameter \(r\). As \(r \rightarrow \infty\), the
negative binomial model behaves arbitrarily similarly to the Poisson
model. We would like to retain this flexibility while assuming a priori
that overdispersion is possible, so we need weakly informative priors.
Given these goals and the domain of the overdispersion parameter
(\(r > 0\)), we chose to focus on Gamma
(\(r \sim G(1, \theta) \equiv r \sim Exp(\theta)\)) and Inverse-Gamma
priors (\(r \sim Inv-Gamma(1, \theta^{-1})\)) using the shape-scale
parameterizations.

Beyond the weakly informative priors, we also tested adding hierarchical
hyper-priors to the scale parameter \(\theta\). First, the
hyperparameter \(\theta\) is either set to a fixed value
\(\eta = 0.1, 1, 10, 100\) or given a hyper-prior
\(\theta \sim Inv Gamma(3, 2 \eta)\) such that \(E(\theta) = \eta\) and
\(Var(\theta) = 1\).

To perform the sensitivity analysis, we re-fit the component models used
for the semi-parametric pregnancy analysis under each of the prior
settings, and collected the posterior distributions of \(r\) and the
LFO-IC for each of the component models used in the stacking.

\begin{figure}

\centering{

\pandocbounded{\includegraphics[keepaspectratio]{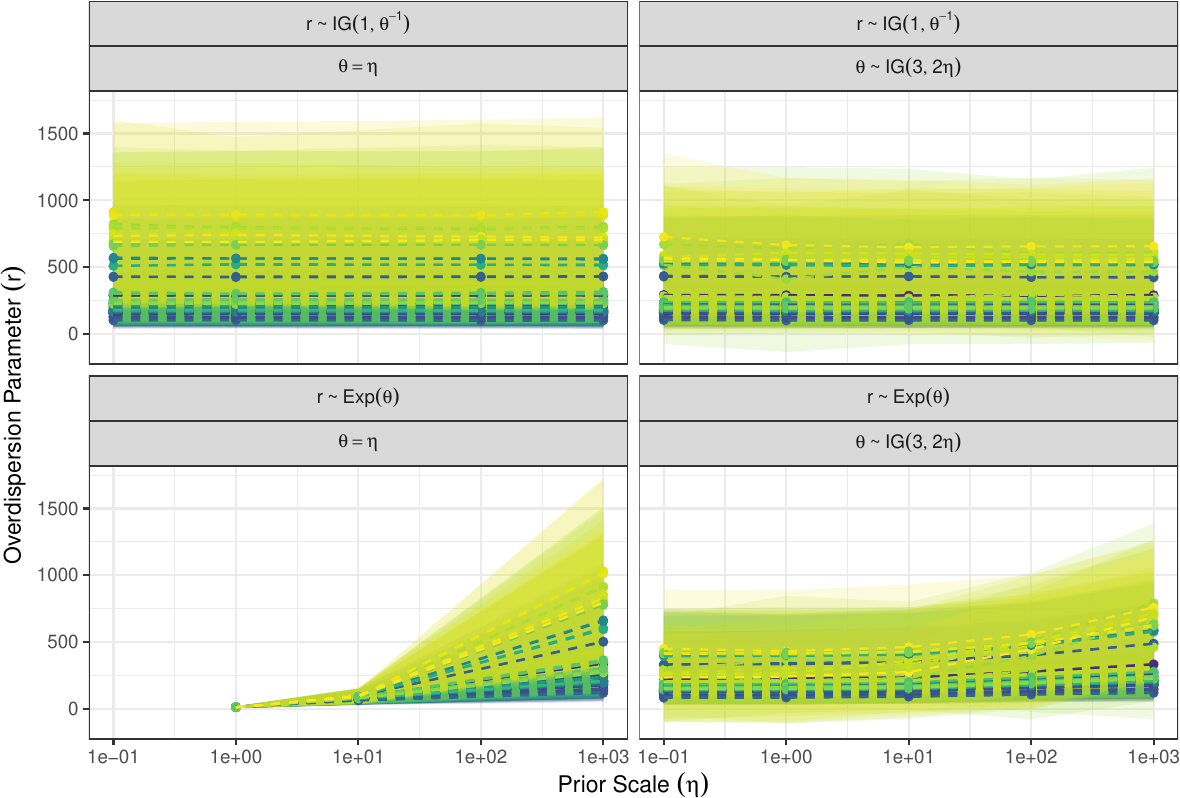}}

}

\caption{\label{fig-overdisp-post}Overdispersion parameter \(r\)'s 95\%
credible interval and posterior mean under the four different priors
with varied prior scale hyperparameter on the x-axis. Each line plot
corresponds to an individual sub-model.}

\end{figure}

In Figure~\ref{fig-overdisp-post}, we compare the overdispersion
parameter's posterior distributions. The left column of
Figure~\ref{fig-overdisp-post} displays posteriors where the
hyperparameter \(\theta\) is held fixed at a pre-specified value
(\(\theta := \eta = 0.1, 1, 10, 100\)). The right column displays
posteriors after assigning a hyper-prior to
\(\theta \sim Inv-Gamma(3, 2 \eta)\) such that \(E(\theta) = \eta\) and
we set \(\eta = 0.1, 1, 10, 100\). The top row sets
\(r \sim Inv-Gamma(1, \theta^{-1})\) while the bottom row sets
\(r \sim Exp(\theta)\). The prior scale hyperparameter's expected value
\(E(\theta) = \eta\) is varied along the x-axis.

From Figure~\ref{fig-overdisp-post}, we can see that while there is
considerable variation in the posterior of \(r\) for each sub-model, the
posteriors stay relatively consistent for the Inverse-Gamma prior.
However, for the exponential priors, the posterior of \(r\) depends on
the prior scale parameter; very strongly if a hyper-prior isn't
specified for \(\theta\). The tendency for the overdispersion
parameter's posterior mean to be between 50 and 200 indicates
mild-overdispersion. Additionally, the insensitivity of the posterior to
the prior scale under the Inverse-Gamma prior suggested we adopt this
prior for the final analysis presented in the application case study, as
the results would not be dependent about the choice of \(\theta\).

We also found that the prior settings mostly do not impact the LFO-ICs
as most LFO-IC estimates for component models overlap even in the case
of the exponential prior on \(r\) with fixed \(\theta\) (lower-right
panel) where the posterior of \(r\) was highly sensitive to the prior
scale.

\end{document}